\documentclass[utf8]{FrontiersinVancouver} 

\usepackage{url,hyperref,lineno,microtype,subcaption}
\usepackage[onehalfspacing]{setspace}
\hypersetup{
    colorlinks=true,
    linkcolor=cyan,
    filecolor=magenta,      
    urlcolor=cyan,
    citecolor=cyan
    }
\usepackage{color}
\newcommand{\modified}[1]{\textcolor{black}{#1}}

\def\keyFont{\fontsize{8}{11}\helveticabold }
\def\firstAuthorLast{Huang {et~al.}} 
\def\Authors{Yixing Huang\,$^{1,2,*}$, Ahmed Gomaa\,$^{1,2}$, Sabine Semrau\,$^{1,2}$, Marlen Haderlein\,$^{1,2}$, Sebastian Lettmaier\,$^{1,2}$, Thomas Weissmann\,$^{1,2}$, Johanna Grigo\,$^{1,2}$, Hassen Ben Tkhayat\,$^{1,3}$, Benjamin Frey\,$^{1,2}$, Udo Gaipl\,$^{1,2}$, Luitpold Distel\,$^{1,2}$, Andreas Maier\,$^{3}$, Rainer Fietkau\,$^{1,2}$, Christoph Bert\,$^{1,2}$, and Florian Putz\,$^{1,2}$}
\def\Address{$^{1}$Department of Radiation Oncology, University Hospital Erlangen, Friedrich-Alexander-Universit\"at Erlangen-N\"urnberg, 91054 Erlangen, Germany \\
$^{2}$Comprehensive Cancer Center Erlangen-EMN (CCC ER-EMN), 91054 Erlangen, Germany\\
$^{3}$Pattern Recognition Lab, Friedrich-Alexander-Universit\"at Erlangen-N\"urnberg, 91058 Erlangen, Germany }
\def\corrAuthor{Yixing Huang}

\def\corrEmail{yixing.yh.huang@fau.de}

\begin{document}
\onecolumn
\firstpage{1}

\title {Benchmarking ChatGPT-4 on ACR Radiation Oncology In-Training (TXIT) Exam and Red Journal Gray Zone Cases: Potentials and Challenges for AI-Assisted  Medical Education and Decision Making in Radiation Oncology} 

\author[\firstAuthorLast ]{\Authors} 
\address{} 
\correspondance{} 

\extraAuth{}

\maketitle

\begin{abstract}


The potential of large language models in medicine for education and decision making purposes has been demonstrated as they achieve decent scores on medical exams such as the United States Medical Licensing Exam (USMLE) and the MedQA exam. In this work, we evaluate the performance of ChatGPT-4 in the specialized field of radiation oncology using the 38th American College of Radiology (ACR) radiation oncology in-training (TXIT) exam and the 2022 Red Journal Gray Zone cases. For the TXIT exam, ChatGPT-3.5 and ChatGPT-4 have achieved the scores of 63.65\% and 74.57\%, respectively, highlighting the advantage of the latest ChatGPT-4 model. Based on the TXIT exam, ChatGPT-4's strong and weak areas in radiation oncology are identified to some extent. Specifically, ChatGPT-4 demonstrates \modified{better} knowledge of statistics, CNS \& eye, pediatrics, biology, and physics \modified{than knowledge of} bone \& soft tissue and gynecology, as per the ACR knowledge domain. Regarding clinical care paths, ChatGPT-4 performs \modified{better} in diagnosis, prognosis, and toxicity \modified{than} brachytherapy and dosimetry. \modified{It lacks proficiency in in-depth details of clinical trials.} For the Gray Zone cases, ChatGPT-4 is able to suggest a personalized treatment approach to each case with high correctness and comprehensiveness. Importantly, it provides novel treatment aspects for many cases, which are not suggested by any human experts. Both evaluations demonstrate the potential of ChatGPT-4 in medical education for the general public and cancer patients, as well as the potential to aid clinical decision-making, while acknowledging its limitations in certain domains. Because of the risk of hallucination, facts provided by ChatGPT always need to be verified.

\tiny
 \keyFont{ \section{Keywords:} ChatGPT, radiation oncology, large language model, artificial intelligence, Gray Zone, decision support, benchmark} 
\end{abstract}

\section{Introduction}

With the recent advances in deep learning techniques such as transformer architectures \cite{vaswani2017attention}, few-shot prompting \cite{brown2020language}, and reasoning \cite{wei2022chain}, large language models (LLMs) have achieved breakthroughs in natural language processing. In these years, many LLMs have been developed and released to the public, including ChatGPT \cite{brown2020language,bubeck2023sparks,openAI2023GPT4}, T5 \cite{raffel2020exploring}, Turing-NLG \cite{smith2022using}, LLaMA \cite{touvron2023llama}, LaMDA \cite{thoppilan2022lamda}, PaLM \cite{chowdhery2022palm} and Alpaca \cite{alpaca}. The capabilities of such LLMs range from simple text-related tasks like language translation and text refinement to complex ones like decision-making \cite{wang2023chat} and programming \cite{chen2021evaluating}, which have already far exceeded public expectation. 

Like other fields, LLMs have also shown great potential in biomedical applications \cite{gu2021domain,lee2020biobert,yunxiang2023chatdoctor}. Several domain-specific language models have been developed such as BioBERT \cite{lee2020biobert}, PubMedBERT \cite{gu2021domain}, and ClinicalBERT \cite{alsentzer2019publicly}. General-domain LLMs with fine-tuning on biomedical data have also achieved impressive results. For example, Med-PaLM \cite{singhal2022large} fine-tuned from PaLM \cite{chowdhery2022palm} has achieved  67.6\% accuracy on the MedQA exam; ChatDoctor \cite{yunxiang2023chatdoctor} fined-tuned from LLaMA \cite{touvron2023llama} using doctor-patient conversation data for more than 700 diseases has achieved 91.25\% accuracy on medication recommendations; HuaTuo \cite{wang2023huatuo} fine-tuned from LLaMA \cite{touvron2023llama} is capable of providing advice on \modified{(traditional and modern)}
 Chinese medicine with safety and usability. In addition to fine-tuning, hybrid models, \modified{which combine LLMs with models of other modalities,} can extend the capabilities of general LLMs. For example, \modified{integrating ChatGPT with advanced imaging networks (e.g. the ChatCAD \cite{wang2023chatcad}) can overcome its limitation in image processing and reach the goal of fully automatic diagnosis from medical images, along with automated medical report generation. }

ChatGPT, being the most successful language model so far, has shown impressive performance in various domains without further fine-tuning, because of the large variety and amount of training data. It has been demonstrated successful in dozens of publicly-available official exams ranging from natural language processing like SAT EBRW reading and writing exams to subject-specific exams such as SAT Math, AP Chemistry and AP Biology exams, as reported in \cite{openAI2023GPT4}. \modified{ChatGPT is capable of improving its performance using reinforcement learning from human feedback (RLHF) \cite{christiano2017deep}.} Because of its excellent performance on multidisciplinary subjects, ChatGPT becomes a very useful tool for diverse users. For domain knowledge in medicine, ChatGPT-3 achieved better than 50\% accuracy across all the exams of the United States Medical Licensing Exam (USMLE) and exceeding 60\% in certain analyses \cite{kung2023performance}; ChatGPT-3.5 has also been reported beneficial for clinical decision support \cite{liu2023using}. Therefore, ChatGPT possesses the potential to enhance medical education for patients and \modified{decision support} for clinicians.

In the field of radiation oncology, deep learning has achieved impressive results in various tasks \cite{sahiner2019deep}, e.g., tumor segmentation \cite{kamnitsas2017efficient,huang2022deep}, lymph node level segmentation \cite{weissmann2023deep}, synthetic CT generation \cite{wang2023improving}, dose distribution estimation \cite{xing2020feasibility}, and treatment prognosis \cite{yang2022computed,hagag2023deep}. With the wide spread of ChatGPT and its \modified{broad} knowledge in medicine, ChatGPT has the potential to be a valuable tool for providing advice to cancer patients and radiation oncologists. Recently, ChatGPT's performance for the specialized domain of radiation oncology physics has been evaluated using a custom-designed exam with 100 questions \cite{holmes2023evaluating}, demonstrating the superiority of ChatGPT-4 to \modified{another LLM Bard}. Nevertheless, the field of radiation oncology covers diverse topics like statistics, biology, and anatomy specific oncology (e.g., gynecologic, gastrointestinal and genitourinary oncology) in addition to physics. To date, the performance of ChatGPT on radiation oncology using standard exams has not been benchmarked yet. Especially, its performance on real clinical cases has not been \modified{fully} investigated. Consequently, the reliability of advice on radiation oncology provided by ChatGPT remains an open question \cite{ebrahimi2023chatgpt}.

In this work, the performance of ChatGPT on the American College of Radiation (ACR) radiation oncology in-training (TXIT) exam and the Red Journal Gray Zone cases is benchmarked. The performance difference between ChatGPT-3.5 and ChatGPT-4 is evaluated. Based on the two evaluations, the confidence zones and blind spots of ChatGPT in radiation oncology are revealed, highlighting its potential to medical education for patients and challenges for aiding clinicians in decision making.

\section{Materials and Methods}

\subsection{Benchmark on the ACR TXIT exam}
\begin{figure}[h]
\raggedright
\includegraphics[width=\linewidth]{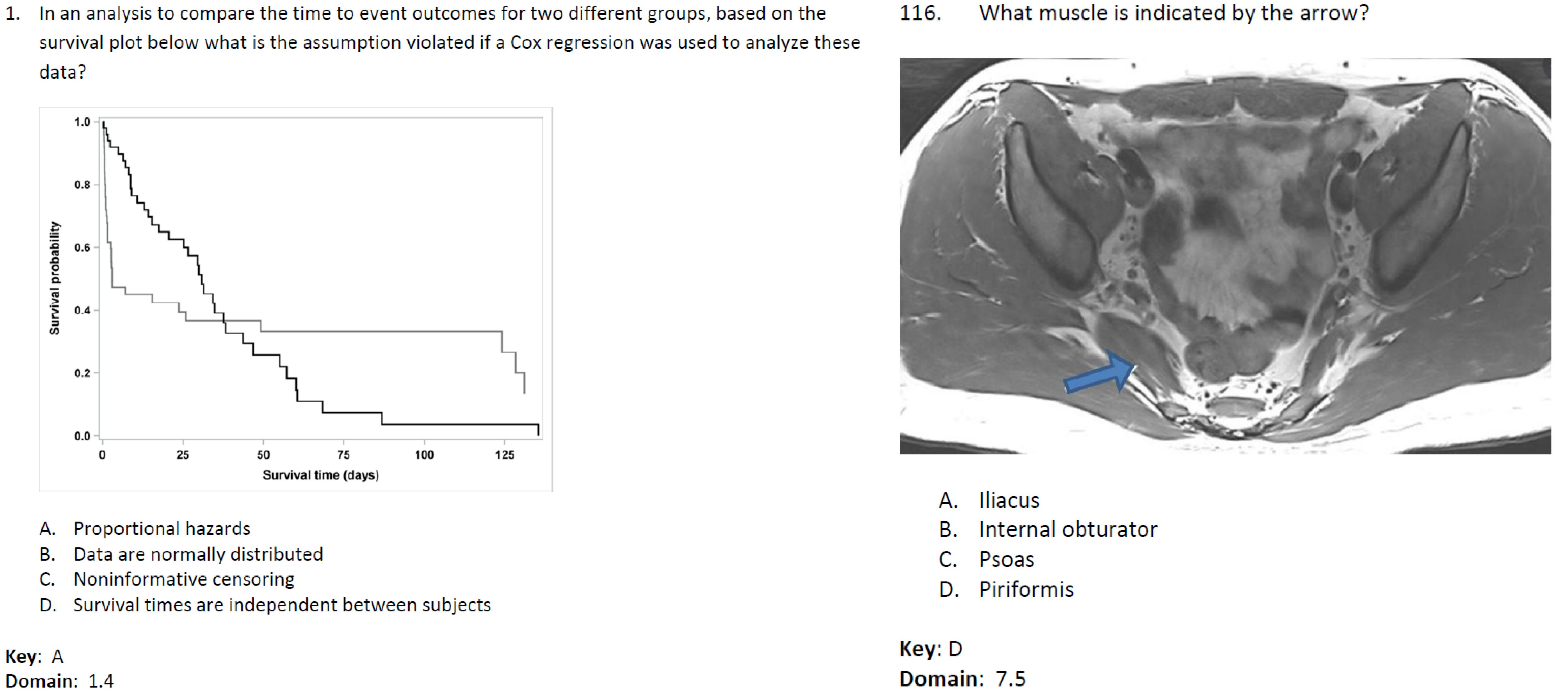}
\caption{Two exemplary questions (Question 1 and Question 116) from the ACR TXIT exam.}
\label{Fig:ExemplaryImageQuestions}
\end{figure}

To benchmark the performance of ChatGPT on radiation oncology, the 38th ACR TXIT exam (2021) is used. The exam sheet is publicly available on the ACR website\footnote{\url{https://www.acr.org/-/media/ACR/Files/DXIT-TXIT/ACR-2021-TXIT-Exam---Assembled.pdf}}. The TXIT exam covers the questions from six primary categories in radiation oncology: diagnosis,  treatment decision, treatment planning, quality assurance, brachytherapy, and toxicity $\&$ management \cite{rogacki2021analysis}. The exam consists of 300 questions in total with 14 questions containing medical images. All questions are multiple choices with single answer. Among the 14 questions with medical images, 7 have a certain text description or list of answers from which the image content or the correct answer can be deduced (e.g., Question 1 as displayed in Fig.\,\ref{Fig:ExemplaryImageQuestions}). However, the other 7 questions \modified{(in particular, Questions 17, 86, 112, 116, 125, 143, and 164)} are impossible to answer from the text information alone without access to the imaging information (e.g., Question 116 as displayed in Fig.\,\ref{Fig:ExemplaryImageQuestions}). Therefore, the later 7 questions are excluded in our evaluation.

In this study, all questions \modified{(questions alone without additional texts)} were entered into ChatGPT. Although no justification was requested in the input prompt, ChatGPT automatically provided certain explanations for its responses.  To maintain consistency, no human feedback was given to ChatGPT. In regard to grading ChatGPT's answers, Question 71 has two correct answers A and B, while ChatGPT-3.5 and ChatGPT-4 both give a single choice answer (A and B, respectively). Therefore, a score of 0.5 is assigned to this question for both ChatGPT-3.5 and ChatGPT-4. In addition, for Question 20 ChatGPT-3.5 suggests A or C depending on whether bone is considered part of the anatomically constrained area; For Question 135, ChatGPT-4 suggests that D may also be a viable option in addition to the correct answer C. For each case, a score of 0.5 is assigned.

\modified{In the initial evaluation of the TXIT exam (where the main findings are based on), the ChatGPT website interfaces for the default ChatGPT-3.5 and the advanced ChatGPT-4 were used. They were accessed in April 2023. While ChatGPT-3.5 is readily available to the public as the standard version, ChatGPT-4 at the time of writing comes with a usage restriction of 25 messages per 3-hour window and is not offered free of charge. In the initial evaluation, a new chat session is recreated after every 5 questions to avoid memory problems. As the chat history influences ChatGPT's responses, for a more fair evaluation, ChatGPT-3.5's and ChatGPT-4's responses on the TXIT exam were assessed again with the ChatGPT API with a temperature parameter of 0.7 on August 10-15th 2023, where a new conversation restarts after each question to avoid the influence of chat history.}

\begin{figure}
\centering
\includegraphics[width=\linewidth]{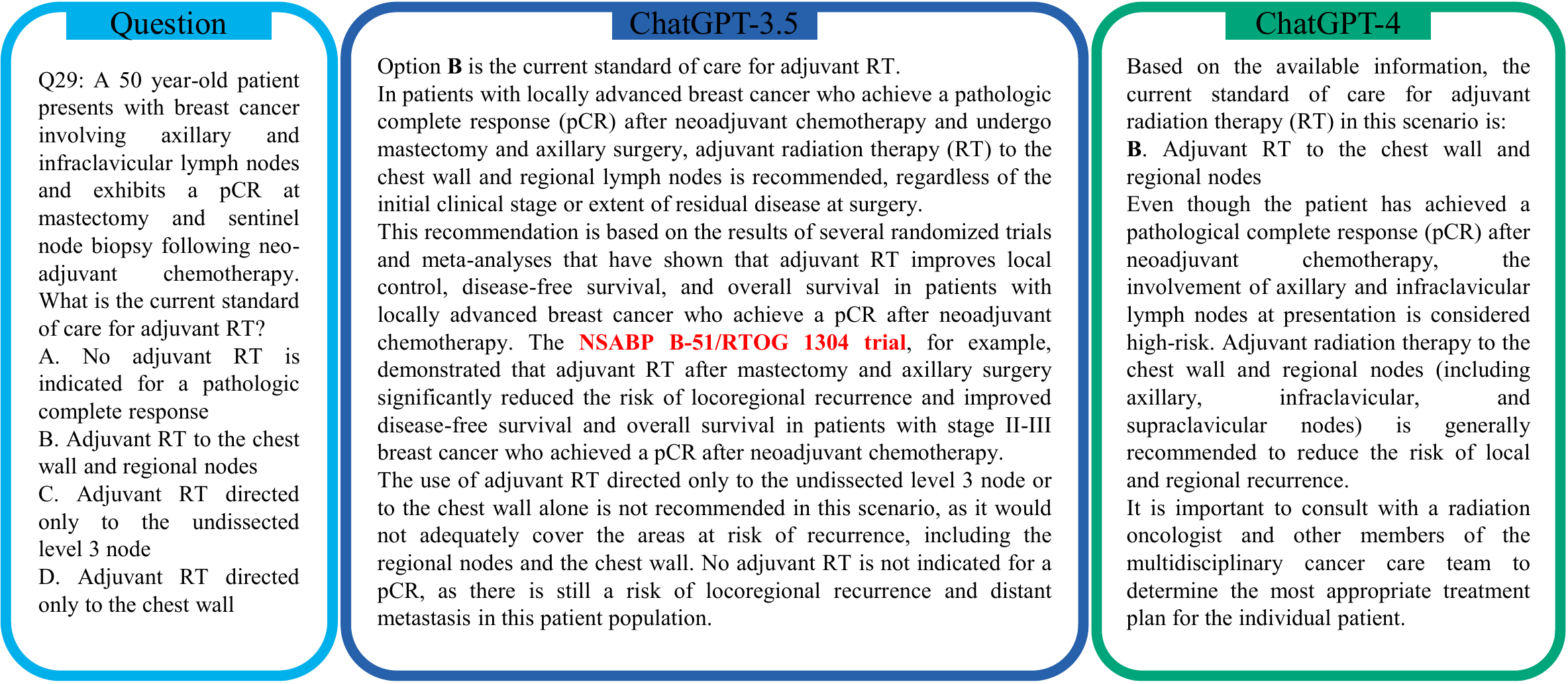}
\caption{An example of question-and-answer using ChatGPT for the ACR TXIT exam. ChatGPT-3.5 and ChatGPT-4 both provide the correct answer. However, ChatGPT-3.5 hallucinates the results of the NSABP B-51/RTOG 1304 trial \cite{Elef2019NRG}, as the final findings are not publicly available yet.}
\label{Fig:QAExample}
\end{figure}

\subsection{Benchmark on the Red Journal Gray Zone cases}
Within the field of radiation oncology as for the whole of medicine, treatment guidelines and available clinical evidence do not always provide a clear recommendation for every clinical case. These difficult clinical situations are referred to as Gray Zone cases \cite{palma2017introducing}, leaving room for differences of opinion and constructive debate in many patient scenarios. 
The vast collection of the Red Journal Gray Zone cases \cite{palma2017introducing} provides ample data that can be used to benchmark the performance of ChatGPT on such real, challenging clinical cases, which traditionally have required highly specialized domain experts and sophisticated clinical reasoning.

In this work, the 2022 collection of the Red Journal Gray Zone cases\footnote{\url{https://www.redjournal.org/content/grayzone}}, in total 15 cases, are used for benchmark. Due to the superior performance of ChatGPT-4 to ChatGPT-3.5 based on the ACR TXIT exam, ChatGPT-4 is used for this evaluation. For each case, ChatGPT-4 is set to a role as an expert radiation oncologist by providing the prompt: ``\textit{You are an expert radiation oncologist from an academic center}", followed by the description of each patient's situation. For diagnostic medical images, only the text captions are provided. Based on the given information for the patient, ChatGPT-4's most favored therapeutic approach and its reasoning for the recommended approach are asked. Afterwards, other experts' recommendations to this case are provided to ChatGPT-4 and the following questions are asked:

$\quad-$\textit{Summarize the recommendations of other experts in short sentences;}

$\quad-$\textit{Which expert's recommendation ChatGPT-4 thinks is the most proper for the patient;}

$\quad-$\textit{ChatGPT-4's initial recommendation is close to which expert's recommendation;}

$\quad-$\textit{Whether ChatGPT-4 will update its initial recommendation after seeing other experts' recommendations.}

For all the Gray Zone cases a clinical expert (senior physician, board-certified radiation oncologist) evaluated the responses of ChatGPT-4 both in a qualitative and semiquantitative manner.
The initial and updated recommendations of ChatGPT-4 were evaluated across 4 dimensions.
First, the correctness of the responses was evaluated on a 4-point Likert scale with the following levels: 4 = ``no mistakes”, 3 = ``mistake in detail aspect not relevant to the validity of the overall recommendation”, 2 = ``mistake in relevant aspect of the recommendation, but recommendation still clinically justifiable”, 1 = “recommendation not clinically justifiable, because of incorrectness”. Moreover, the comprehensiveness of the recommendation was also evaluated on a 4-point Likert scale with the following levels: 4 = ``recommendation covers all relevant clinical aspects”, 3 = ``recommendation is missing some detail information, e.g., in regard to radiotherapy dose or target volume”, 2 = ``recommendation is missing relevant aspect, but overall recommendation is still clinically justifiable”, 1 = ``recommendation not clinically justifiable, because of incompleteness”. Finally, novel valuable aspects in ChatGPT-4’s response, not present in the real clinical experts’ recommendations, as well as hallucinations were rated in a binary manner (``present” vs. ``not present”). \modified{Hallucinations are responses generated by LLMs in a convincing appearance but actually are incorrect statements \cite{openai2023system}.} Ratings for initial and revised recommendations were tested for difference using a paired Wilcoxon test. Aside from the initial and the final recommendation, all other responses by ChatGPT-4 were evaluated in a qualitative manner.

\section{Results}
\subsection{\textbf{Results on the ACR EXIT exam}}
\subsubsection{Overall performance difference between ChatGPT-3.5 and ChatGPT-4}

Across the total of 293 questions, ChatGPT-3.5 and ChatGPT-4 attain accuracies of 63.14\% and 74.06\% \modified{respectively in our initial assessment via the website interface, both surpassing the standard pass rate of 60\%. In the 5-time repeated assessment via the ChatGPT API, ChatGPT-3.5 and ChatGPT-4 have achieved average accuracies of 62.05\% $\pm$ 1.13\% and 78.77\% $\pm$ 0.95\%, respectively.} The advanced ChatGPT-4 version exhibits a \modified{16.72\%} increase in accuracy compared to the standard ChatGPT-3.5, illustrating its superior performance in the field of radiation oncology. Out of all the questions that are answered incorrectly, ChatGPT-3.5 and ChatGPT-4 are both incorrect for 51 of them. Additionally, ChatGPT-3.5 is incorrect for 57 questions that are correctly answered by ChatGPT-4, while ChatGPT-4 is incorrect for 25 questions that are correctly answered by ChatGPT-3.5. Please find the answers to all the questions in the supplementary material. A copy of answers is available via GitHub\footnote{\href{https://github.com/YixingHuang/ChatGPT-Benchmark-on-Radiation-Oncology}{https://github.com/YixingHuang/ChatGPT-Benchmark-on-Radiation-Oncology}}.

Figure 2 illustrates an exemplary instance of question-answering, where both ChatGPT-3.5 and ChatGPT-4 successfully deliver accurate responses. In general, ChatGPT-3.5 provides longer answers with a quicker generation speed than ChatGPT-4. In this example, ChatGPT-3.5 refers to the NSABP B-51/RTOG 1304 trial \cite{Elef2019NRG} for its justification of the answer. Instead, ChatGPT-4 typically provides an answer with a shorter explanation. More frequently, ChatGPT-4 will include a cautionary message to prevent users from being inadvertently led towards potential health hazards, e.g., ``\textit{It is important to consult with a radiation oncologist and other members of the multidisciplinary cancer care team to determine the most appropriate treatment plan for the individual patient}" at the end of the answer to this exemplary question.

\subsubsection{Domain dependent performance}

\begin{figure}[t]
\includegraphics[width=\linewidth]{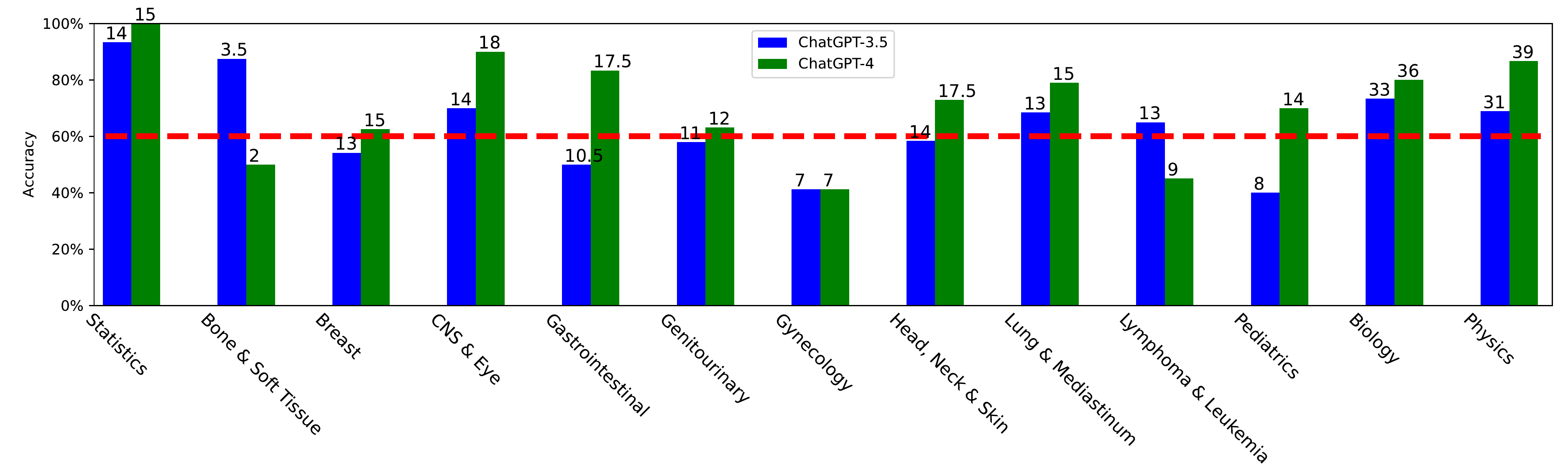}
\caption{The accuracy distribution for ChatGPT-3.5 and ChatGPT-4 depending on the question domain. The absolute number of correct answers for each domain is marked at the top of each bar. The domain number 1-13 correspond to statistics, bone $\&$ soft tissue, breast, CNS $\&$ eye, gastrointestinal, genitourinary, gynecology, head \& neck $\&$ skin, lung $\&$ mediastinum, lymphoma $\&$ leukemia, pediatrics, biology, and physics, respectively. The X-axis labels are shifted to save space.}
\label{Fig:domainAccuracy}
\end{figure}

All the questions of the ACR TXIT exam belong to 13 major knowledge domains, according to the TXIT table of specifications\footnote{\url{https://www.acr.org/-/media/ACR/Files/DXIT-TXIT/ACR-TXIT---Table-of-Specifications.pdf}}. The 13 domains are statistics, bone $\&$ soft tissue, breast, central nervous system (CNS) $\&$ eye, gastrointestinal, genitourinary, gynecology, head \& neck $\&$ skin, lung $\&$ mediastinum, lymphoma $\&$ leukemia, pediatrics, biology, and physics. The accuracies achieved by ChatGPT-3.5 and ChatGPT-4 for different domains are displayed in Fig.\,\ref{Fig:domainAccuracy}. Considering a 60\% threshold, ChatGPT-3.5 only obtains 54.17\%, 50\%, 57.89\%, 41.18\%, 58.33\%, and 40\% for breast, gastrointestinal, genitourinary, gynecology, head \& neck \& skin and pediatrics, respectively. Notably, its accuracy for gynecology and pediatrics is only around 40\%. In contrast, for statistics, CNS \& eye, and biology, ChatGPT-3.5 achieves accuracies higher than 70\%. 

ChatGPT-4 attains a worse accuracy than ChatGPT-3.5 for bone \& soft tissue and lymphoma \& leukemia with accuracies of 50\% and 45\%, respectively. Note that there are only 4 valid questions for bone \& soft tissue, and ChatGPT-4 answers 2 questions incorrectly (Question 18: \textit{What postoperative RT dose is recommended for a high grade malignant peripheral nerve sheath tumor of the upper extremity following R1 resection?} Question 19: \textit{What is the recommended preoperative GTV to CTV target volume expansion for an 8.5 cm high grade myxofibrosarcoma of the vastus lateralis muscle?}). ChatGPT-4 gets the same bad performance (41.18\%) on gynecology. ChatGPT-4 outperforms ChatGPT-3.5 in all other domains, with particularly impressive results of 100\% accuracy for statistics, 90\% accuracy for CNS \& eye, 83.33\% accuracy for gastrointestinal, and 86.67\% accuracy for physics.

\subsubsection{Clinical care related performance}
Out of all the questions, the majority (totalling 190) are related to clinical care. Thus, beyond the standard ACR TXIT domain-dependent performance evaluation, these 190 questions---excluding those related to statistics, biology, and physics---are further classified into the following clinical care path-related categories: diagnosis, treatment decision, treatment planning, prognosis, toxicity, and brachytherapy. The prognosis category covers questions related to patient survival and tumor recurrence rate/risk, while the toxicity category covers questions related to side effects of treatment. Since dosimetry plays a crucial role in radiation therapy, all the dose-related questions are also grouped into one common category. Note that the categorization is not exclusive, which means that one question might belong to more than one category.

\begin{figure}[t]
\centering
\includegraphics[width=\linewidth]{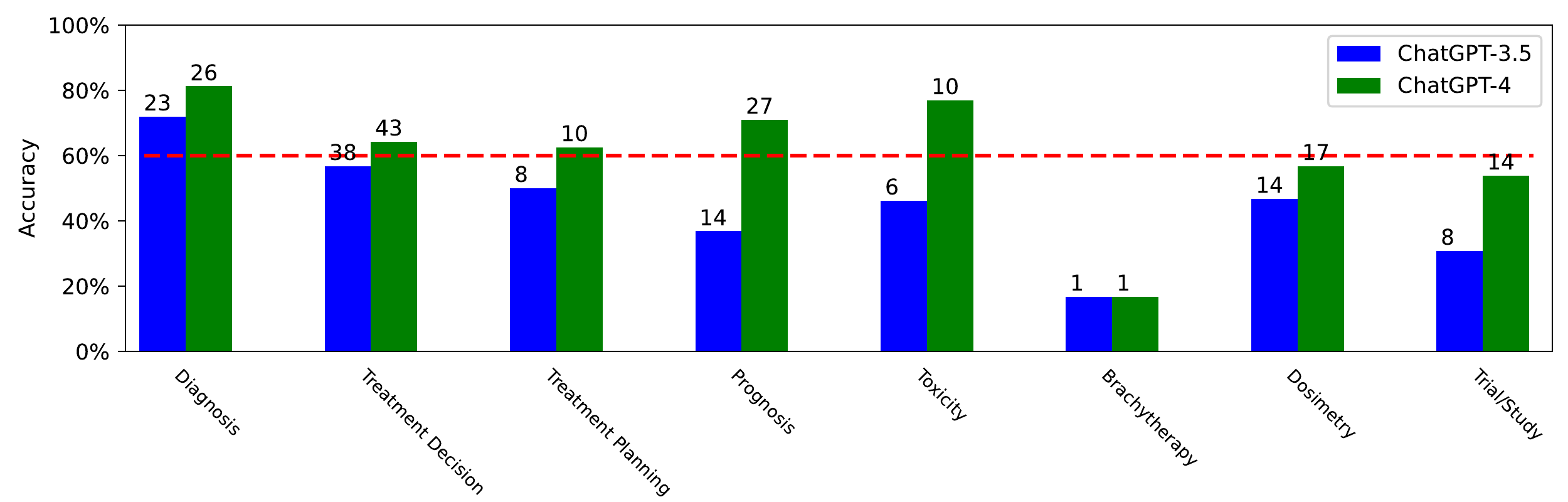}
\caption{The accuracy distribution for ChatGPT-3.5 and ChatGPT-4 depending on the clinical care category. The absolute number of correct answers for each domain is marked at the top of each bar. The category number 1-8 correspond to diagnosis, treatment decision, treatment planning, prognosis, toxicity, brachytheraphy, dosimetry, and trial/study, respectively. The X-axis labels are shifted to save space.}
\label{Fig:calinicalCareAccuracy}
\end{figure}

The accuracy distribution of ChatGPT-3.5 and ChatGPT-4 depending on the clinical care categories is displayed in Fig.\,\ref{Fig:calinicalCareAccuracy}. 
Considering the 60\% threshold, ChatGPT-3.5 passes this threshold only for diagnosis with a high accuracy of 71.88\%. However, the accuracies for all the remaining categories are 56.72\%, 50\%, 36.84\%, 46.15\%, 16.67\%, and 46.67\%, respectively, all lower than 60\%. In comparison, ChatGPT-4 passes the threshold for diagnosis, treatment decision, treatment planning and toxicity with accuracies of 81.25\%, 62.5\%, 71.05\%, and 76.92\%, respectively. But its accuracies for brachytherapy and dosimetry are lower than 60\%, which are 16.67\% and 56.67\%, respectively. Out of all the categories, both ChatGPT-3.5 and ChatGPT-4 exhibit similarly unsatisfactory performance on brachytherapy, as they are only capable of correctly answering the same specific single question from the total of six presented. Among other categories, ChatGPT-4 exhibits superior performance compared to ChatGPT-3.5, particularly in the areas of prognosis and toxicity, where ChatGPT-4 surpasses its predecessor by 30\%.


\subsubsection{Performance on clinical trials}
Among all the questions, many questions are based on certain clinical trials (e.g., the Stockholm III \cite{erlandsson2017optimal}, the CRITICS randomized trial \cite{cats2018chemotherapy}, the PORTEC-3 trial \cite{de2019adjuvant}, the German rectal study \cite{kirchheiner2016dose}, and the ORIOLE phase 2 randomized clinical trial \cite{phillips2020outcomes}) 
or guidelines (e.g., the 8th AJCC cancer staging manual \cite{amin2017eighth}). Such questions are also grouped into a category called trial/study, and the accuracies of ChatGPT-3.5 and ChatGPT-4 are displayed in Fig.\,\ref{Fig:calinicalCareAccuracy} as well.

ChatGPT-3.5 and ChatGPT-4 obtain the accuracies of 30.77\% and 53.85\% on the trial/study related questions, respectively, both of which are lower than 60\%. ChatGPT-4 achieves 23\% higher accuracy than ChatGPT-3. If we ask ChatGPT-3.5 and ChatGPT-4 whether they know a certain trial, e.g., the PORTEC-3 trial \cite{de2019adjuvant}, both of them will provide a positive answer ``Yes, I am familiar with the PORTEC-3 trial" and provide a short summary of the mentioned trial. This suggests that both ChatGPT-3.5 and ChatGPT-4 have encountered such trials and studies in their training data. However, there is still a significant risk of them providing inaccurate answers. For example, for Question 107 (``\textit{In the subset analysis of PORTEC-3 trial, patients with which histology MOST benefited from the addition of chemotherapy to RT? A. Endometrioid B. Carcinosarcoma C. Clear cell D. Serous}"), ChatGPT-3.5 answers A endometrioid, while ChatGPT-4 answers B carcinosarcoma, both of which are incorrect. If we copy and paste the summary/abstract of the PORTEC-3 trial \cite{de2019adjuvant} into the conversation, i.e., leverage ChatGPT's in-context learning capabilities, and ask ChatGPT the question again, both ChatGPT-3.5 and ChatGPT-4 can provide the correct answer D based on the given summary.
 
 It is worth noting that in the example of Fig.\,\ref{Fig:QAExample}, the NSABP B-51/RTOG 1304 has not been fully published yet and only a meeting update on the course of the study without any definitive results was included in the short abstract of \cite{Elef2019NRG}. Therefore, ChatGPT-3.5 is hallucinating the results here in Fig.\,\ref{Fig:QAExample}.

%
%
%
%
%
%
%
%

\subsection{\textbf{Results for clinical decision making in the Gray Zone cases}}

\begin{table}[t]
\caption{The performance of ChatGPT-4's initial recommendations and revised recommendations on the Gray Zone cases.}
\label{Tab:grayZone}
\centering
\begin{tiny}
\begin{tabular}{p{2.0cm}p{0.5cm}p{0.5cm}p{0.5cm}p{0.5cm}p{0.5cm}p{0.5cm}p{0.5cm}p{0.5cm}p{0.5cm}p{0.75cm}p{0.5cm}p{0.5cm}p{0.75cm}p{0.5cm}p{0.5cm}}
\hline
Case ID & 1 & 2 & 3 & 4 & 5 & 6 & 7 & 8 & 9 & 10 & 11 & 12 & 13 & 14 & 15\\
\hline
\multicolumn{15}{l}{\textbf{Distribution of votes for the Gray Zone clinical expert recommendations:}}\\
Expert 1 &61.54 & 20 & 5.56 & 7.14 & 71.43 & 60 & 40 & 8 & 29.41 & 60 & 37.5 & 62.5 & 0 & 25 & 16.67 \\
Expert 2 & 15.38 & 26.67 & 0 & 57.14 & 0 & 40 & 20 & 0 & 52.94 &30 &25 & 12.5 & 100 & 50 & 50\\
Expert 3 &0 & 33.33 & 55.56 & 35.71 & 14.29 & 0 & 40 & 32 & 5.88 & 10 & 25 &12.5 & 0 & 25 & 33.33\\
Expert 4 &0 & 20 & 38.89 & - & 0 & - & - & 20 & 11.76 & - & 12.5 & 12.5 & - & - & -\\
Expert 5 & 23.08 & - & - & - & 14.29 & - & - & 40 & - & - & - & - &- &- &-\\
\hline
\multicolumn{15}{l}{\textbf{GPT-4's self-assessment:}}\\
Closest & E3 & E2 & E1 & E1 & E4 & E2 &E1 & E3 & E3 &E1+E2 & E3 &E1 & E2+E3 & E2 & E2+E3\\
Favourite &E3 & E3 & E4 &E1 &E2 & E2 & E2 & E2 & E2 & E1+E2 & E3 & E2 &E1 & E2 & E2\\
\hline 
\multicolumn{15}{l}{\textbf{Senior physician's assessment:}}\\
\multicolumn{15}{l}{Initial recommendation}\\
Correctness & 4 & 4 & 3 & 4 & 4 & 4 &3 & 2 & 4 &3 & 4 &3 & 4 & 3 & 4\\
Comprehensi.&3 & 4 & 3 &2 &3 & 2 & 4 & 2 & 4 & 4 & 3 & 3 &2 & 4 & 4\\
Novel aspects&Yes & Yes & No &Yes &No & Yes & Yes & Yes & No &Yes & Yes& Yes &Yes & Yes & Yes\\
Hallucination &No & No & No &No &No & No & No & Yes & No & Yes & No & No &No & No & No\\
\multicolumn{15}{l}{Revised recommendation}\\
Correctness &4 & 4 & 4 &4 &4 & 4 & 4 & 4 & 4 & 4 & 4 & 4 &4 & 4 & 4\\
Comprehensi. &3& 4 & 3 &4 &4 & 4 & 3 & 4 & 4 &  4 & 3 & 4 &3 & 4 & 4\\
Novel aspects &Yes & No & No &Yes &No & Yes & No& Yes& No & No & No & No &No & No & Yes\\
Hallucination &No &No & No &No &No & No & No& No & No & No & No & No &No & No & No\\
\hline 
\end{tabular}
\raggedright

Closest: ChatGPT-4's initial recommendation is closest to which expert's recommendation.

Favourite: Which expert's recommendation is the most proper for the patient.

Comprehensi. = Comprehensiveness

\end{tiny}
\end{table}

\begin{figure}[h!]
\centering
\includegraphics[width=0.82\linewidth]{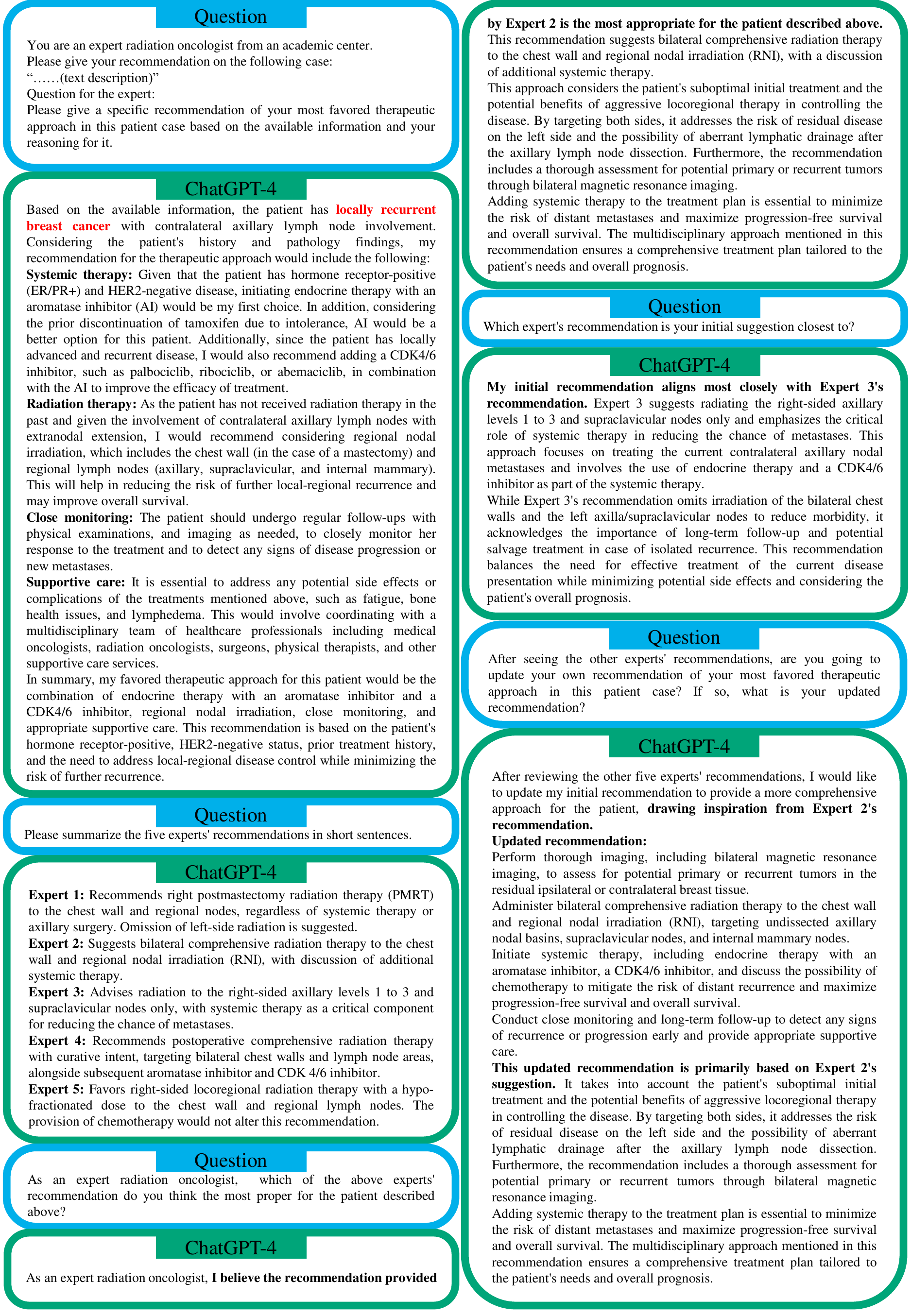}
\caption{An example of ChatGPT-4's recommendation for the Gray Zone case $\#$8 \cite{al2022viewpoint}: A viewpoint on isolated contralateral axillary lymph node involvement by breast cancer: regional recurrence or distant metastasis? Note that the local recurrence statement in ChatGPT-4's summary is incorrect.}
\label{Fig:GrayZoneExemple8}
\end{figure}

\subsubsection{An exemplary Gray Zone case}
ChatGPT-4's recommendations for one exemplary Gray Zone case \cite{al2022viewpoint} are displayed in Fig.\,\ref{Fig:GrayZoneExemple8}. In this example, a 55-year-old woman was treated in the past for mixed invasive lobular and ductal carcinoma of the left breast developed contralateral nodal recurrence after an interval of 10 years. More details about this patient can be found in \cite{al2022viewpoint} and its detailed responses to other cases are available in our \href{https://github.com/YixingHuang/ChatGPT-Benchmark-on-Radiation-Oncology}{GitHub repository}. ChatGPT-4 proposes a combination of endocrine therapy \cite{tremont2017endocrine} with an aromatase inhibitor and a CDK4/6 inhibitor, regional nodal irradiation, close monitoring and appropriate supportive care. ChatGPT-4 makes such a recommendation based on the patient's genotype and historic treatment. In addition, ChatGPT-4 provides a concise summary of the five experts' recommendations, as displayed in Fig.\,\ref{Fig:GrayZoneExemple8}. ChatGPT-4 thinks that its initial recommendation aligns most closely with Expert 3's recommendation because both suggest focusing on treating the current contralateral axillary nodal metastases and involve the use of endocrine therapy and a CDK2/6 inhibitor as part of the systemic therapy. Nevertheless, ChatGPT-4 favors Expert 2's recommendation instead of Expert 3's since Expert 2's recommendation considers the patient's suboptimal initial treatment and the potential benefits of aggressive locoregional therapy in controlling the disease. Therefore, after seeing all five experts' opinions, ChatGPT-4 tends to update its recommendation, ``drawing inspiration from Expert 2's recommendation". This exemplary case demonstrates the potential of ChatGPT-4 in assisting decision making for intricate Gray Zone cases.

\subsubsection{Overall performance based on ChatGPT-4's self-assessment}
For the Gray Zone cases, the recommendations provided by clinical experts for each case are voted by other expert raters and published on the \href{https://www.redjournal.org/content/grayzone}{Red Journal website}. The distribution of votes for the 15 cases in the 2022 collection is displayed in Tab.\,\ref{Tab:grayZone}. For instance, for the first case \cite{tchelebi2022sowing}, Expert 1, Expert 2, and Expert 3 received 61.54\%, 15.38\%, and 23.08\% of votes, respectively, out of a total of 13 votes, while Expert 3 and Expert 4 received no votes. A total of 59 expert recommendations were evaluated for the 15 Gray Zone cases, resulting in an average vote of 25.42\% (15/59) for each expert. ChatGPT-4's initial recommendation for each case typically shares common points with a specific expert, which has two implications: (a) ChatGPT-4's recommendation is comparable to that of a human expert, and (b) ChatGPT-4's recommendation provides complementary information to that of other individual experts. For certain cases (Case $\#$10 \cite{scarpelli2022exploring}, Case $\#$13 \cite{prpic2022radiation}, and Case $\#$15 \cite{johnson2021synopsis}), ChatGPT-4's recommendation covers points from two experts, indicating that ChatGPT-4's recommendation is more comprehensive. If we consider the closest expert vote to ChatGPT-4's recommendation as an approximate evaluation metric, ChatGPT-4's recommendation would receive an average vote of 28.76\%. Similarly, ChatGPT-4's preferred recommendation from other experts receives an average vote of 24.99\%. Both values (28.76\% and 24.99\%) are close to the experts' average vote of 25.42\%.



\modified{In the evaluation of ChatGPT-4's recommendations for the Gray Zone cases,} ChatGPT-4 consistently highlights the significance of a multidisciplinary team in seeking an individualized treatment plan that is balanced or comprehensive. Its preferred recommendations typically take into account the patient's historic treatment responses, survival prognosis, and potential risks and toxicity. Additionally, it respects the personal preferences and priorities of the patient.

\subsubsection{Clinical evaluation of ChatGPT4’s recommendations}
The initial recommendations of ChatGPT-4 generally show a similar structure: After a summary of the clinical case, ChatGPT-4 provided its recommendations for the patient vignette; At the end of the recommendations, more generic, universally agreed statements e.g., on the value of interdisciplinary discussion, post-treatment follow-up and supportive treatment, were typically observed (e.g., Fig.\,\ref{Fig:GrayZoneExemple8}). 
In the initial case summary, ChatGPT-4 typically was able to capture the relevant aspect of the patient case in a correct manner. Errors were rarely observed in the case summary, but one example can be seen in the exemplary case (Case $\#$8 \cite{al2022viewpoint} in Fig.\,\ref{Fig:GrayZoneExemple8}). In Fig.\,\ref{Fig:GrayZoneExemple8}, the patient case is described as ``\textit{locally recurrent breast cancer with contralateral axillary lymph node involvement}." However, this is not fully correct, as the patient had no local recurrence but only contralateral axillary lymph node metastasis. As the information on local recurrence is not present in the case vignette, this is also rated as hallucination (Tab.\,\ref{Tab:grayZone}).

Generally, the initial recommendations of ChatGPT-4 showed a surprising amount of correctness and comprehensiveness. On a scale of 1-4, mean correctness of the 15 initial ChatGPT-4 recommendations is 3.5 and mean comprehensiveness is 3.1. Importantly, no recommendation is rated with the lowest score of 1, meaning that all ChatGPT-4 recommendations are seen as clinically justifiable. Hallucinations are \modified{common} with 13.3\% (2/15) of recommendations. 

Interestingly, ChatGPT-4 show some signs of clinical reasoning, in which specific features of the case vignette are integrated into the decision making beyond standard clinical algorithms or in which recommendations are obtained by multi-step reasoning. For example in Fig.\,\ref{Fig:GrayZoneExemple8}, ChatGPT-4 recommends an aromatase inhibitor (AI) by referring to the patient’s prior intolerance to Tamoxifen (“[…] \textit{considering the prior discontinuation of tamoxifen due to intolerance, AI would be a better option for this patient.} […]”). A quite impressive example for multi-step clinical reasoning can be found in the recommendation for Case $\#$9 \cite{berghen2022postoperative} (\href{https://github.com/YixingHuang/ChatGPT-Benchmark-on-Radiation-Oncology}{Supplementary material}) on a recurrent prostate cancer case, in which an isolated supraclavicular lymph node showed PSMA-uptake: ``\textit{The 18F-PSMA PET/CT shows no evidence of locoregional recurrence but identifies a hypermetabolic lymph node in the supraclavicular region on the left side. This finding could represent metastatic involvement, although it is unusual for prostate cancer to metastasize to this location without involving pelvic lymph nodes first. […] it is important to consider a biopsy of the suspicious supraclavicular lymph node to confirm its nature and guide further management, which might include directed radiation therapy to the supraclavicular region if it is confirmed to be metastatic}”. 

Finally, it is interesting to note that we also observe novel valuable aspects in ChatGPT-4’s recommendations that are not present in the real clinical experts’ responses with a frequency of 80.0\% (12/15). These include the recommendation for considering tumor-treating field in a patient with grade 2 glioma (Case $\#$10 \cite{hottinger2016tumor}), the recommendation to incorporate potential drug interaction when choosing concurrent chemotherapy in a liver transplant patient (Case $\#$13 \cite{prpic2022radiation}), the recommendation for genomic profiling to find targetable molecular alterations for systemic therapy in a patient case with oligometastatic breast cancer (Case $\#$2 \cite{goodman2022one}) as well as the consideration for additional immunotherapy in a patient with anal cancer (Case $\#$1 \cite{tchelebi2022sowing}).

In the subsequent conversation with ChatGPT-4 following its initial recommendation, the LLM generally was able to summarize the relevant aspects of the clinical experts’ recommendation well. In the updated ChatGPT-4 recommendation following in-context learning with the clinical expert recommendations, we observe a significant increase in correctness (mean, 4.0 vs. 3.5, paired p = 0.020), comprehensiveness (mean, 3.7 vs. 3.1, p = 0.046), no hallucinations (0.0\% vs. 13.3\%) but also reduced novel aspects (33.3\% vs. 80.0\%). Interestingly, in the revised ChatGPT-4 recommendation particularly valuable aspects of the clinical experts' recommendation are incorporated. This e.g., includes the recommendation for neoadjuvant chemotherapy in a patient with metastatic anal cancer that received prior prostate brachytherapy (Case $\#$1 \cite{tchelebi2022sowing}) to select, if the patient will benefit from locoregional treatment as well as the recommendation to perform bilateral chest wall MRI in a patient case with contralateral supraclavicular lymph node recurrence from breast cancer (Case $\#$8 \cite{al2022viewpoint} in Fig.\,\ref{Fig:GrayZoneExemple8}).

\section{Discussion}

 \begin{figure}
 \centering
 \includegraphics[width=1.0\linewidth]{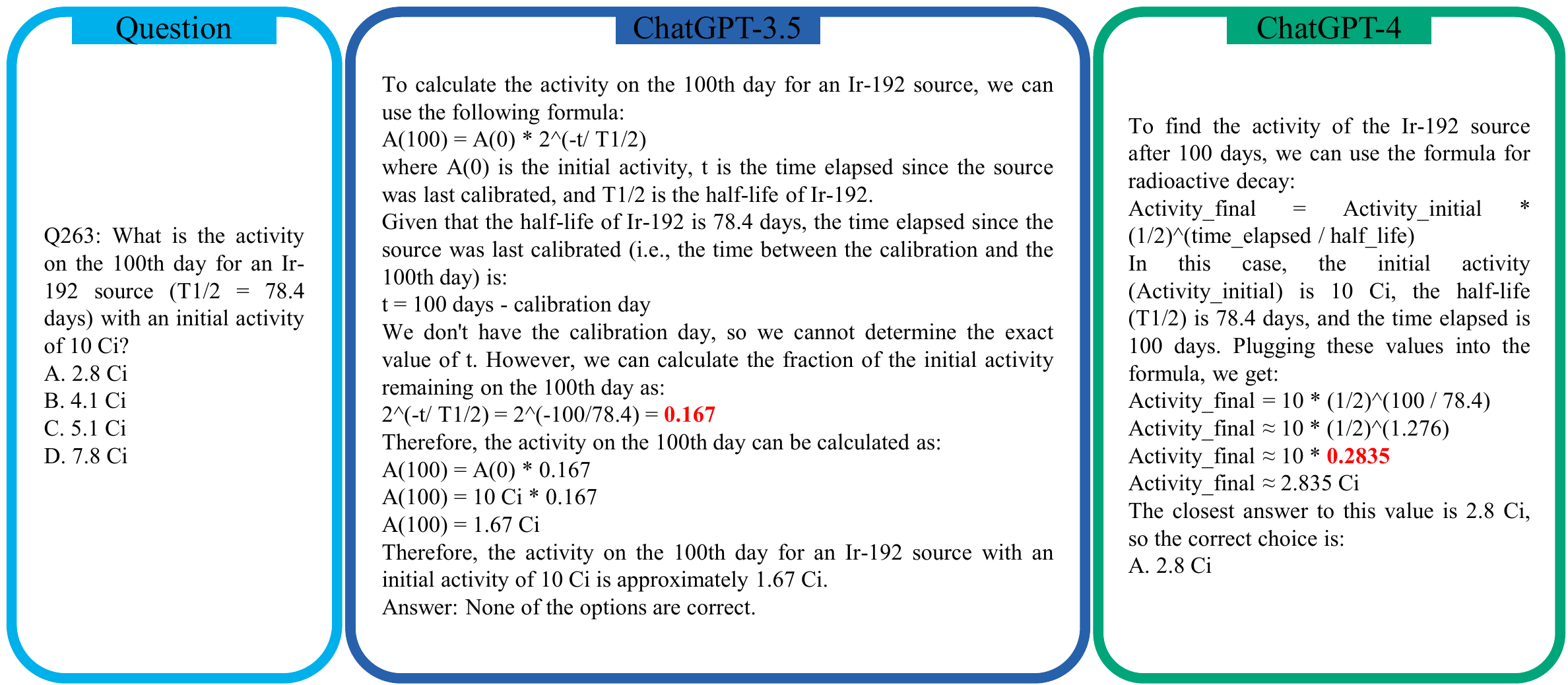}
 \caption{The incorrect exponential decay values calculated by ChatGPT-3.5 and ChatGPT-4 for radioactive decay in Question 263 of the ACR TXIT exam.}
 \label{Fig:Q263Math}
 \end{figure}
 
\subsection{Potential for medical education in radiation oncology}
\modified{Both ChatGPT-3.5 and ChatGPT-4 have demonstrated a certain level of proficiency in grasping fundamental concepts of radiation oncology.} For instance, both versions of ChatGPT are proficient in identifying common types of cancer and have a certain awareness of clinical trials and studies. In the evaluation of Gray Zone cases, ChatGPT-4 provides reasonable explanations to the recommended treatment approach and receives high ratings for correctness and comprehensiveness. \modified{As such, ChatGPT possesses the potential to offer medical education on radiation oncology to the general public and cancer patients,} promoting radiation oncology education into a new stage \cite{oertel2022digital}, when the risk of content hallucination is accounted for by proper verification.

\subsection{Potential to assist in clinical decision making}
ChatGPT-4's decent performance on the topics of diagnosis, treatment decision, treatment planning, prognosis and toxicity (Fig.\,\ref{Fig:calinicalCareAccuracy}) and its reasonable responses on the Gray Zone cases (Fig.\,\ref{Fig:GrayZoneExemple8} and Tab.\,\ref{Tab:grayZone}) indicate its potential to assist in clinical decision making. As a single human expert may fail to consider all aspects of an intricate Gray Zone case, ChatGPT-4's recommendation can provide valuable complementary information in certain cases, potentially leading to a more comprehensive treatment approach. Especially, ChatGPT is capable of suggesting novel treatment ideas (novel aspects in Tab.\,\ref{Tab:grayZone}), for example, using tumor-treating field \cite{hottinger2016tumor} for the patient case with gliomas \cite{scarpelli2022exploring}, which was not suggested by any of the human experts. Therefore, such general artificial intelligence like ChatGPT can in turn improve human decision-making by increasing novelty \cite{shin2023superhuman}.

\subsection{Challenges in clinical decision making for certain topics}

While both versions of ChatGPT exhibit a grasp of essential concepts in radiation oncology, their knowledge is limited or superficial when it comes to certain topics such as gynecology, brachytherapy, dosimetry, and clinical trials, based on the TXIT exam. Consequently, these areas may elicit a relatively high rate of false responses when queried in-depth. In the evaluation of the Gray Zone cases, the recommendations closest to ChatGPT4's response or its favored expert recommendations have received low votes from the rators for some cases, like Case 1, Case 4, Case 5, and Case 13 in Tab.\,\ref{Tab:grayZone}. Therefore, despite of its potential, ChatGPT still has certain limitations in clinical decision making.


\subsection{Addressing the risk of hallucination}
LLMs can hallucinate facts \cite{azamfirei2023large,guerreiro2023hallucinations}, when generating responses, which becomes a widely-known limitation. Especially, for the field of medicine, it is critical that the provided information is correct. We mostly observed hallucinations in the context of clinical trials and citations. For example in this evaluation, ChatGPT-3.5 cited results of the NSABP B-51/RTOG 1204 study (Fig.\,\ref{Fig:QAExample}), which have not been published yet; ChatGPT-4 falsely added the local recurrence in the summary of Case $\#$8 in Fig.\,\ref{Fig:GrayZoneExemple8}. The tendency of LLMs for hallucinations may be particularly problematic as they may be missed by the less proficient reader, since the hallucinations frequently appear very plausible in the context of the text and are phrased in a convincing manner.
Because of the risk of content hallucination, answers and recommendations by ChatGPT always need to be verified. Potential solutions to reduce the risk of hallucination include in-context learning as well as model fine-tuning on medical studies and guidelines.

\subsubsection{Responses on math need cross-check}
Regarding simple math questions like calculating the mode (Question 6), mean (Question 7), and median (Question 8) of a sequence, both ChatGPT-3.5 and ChatGPT-4.0 are able to solve them correctly. However, for exponential/radioactive decay, although ChatGPT-3.5 and ChatGPT-4 both know the right mathematical expression, they nevertheless are likely to provide an incorrect value. For example, for the radioactive decay calculation in Question 263, both ChatGPT-3.5 and ChatGPT-4 could not obtain the correct value for $2^{(-100/78.4)} = 2^{(-1.2755)}  = 0.413$, as displayed in Fig.\,\ref{Fig:Q263Math}. We observed that if this question is asked multiple times in new conversations, each time a different value will be generated by ChatGPT-3.5 and ChatGPT-4. If ChatGPT-4 is asked to calculate the intermediate step $(1/2)^{(100/78.4)}$ specifically, it is able to provide a more accurate answer of 0.4129 and hence updates its previous answer to the correct answer B 4.1 Ci, while ChatGPT-3.5 still fails with an inaccurate value 0.2456. It is worth noting that at the time of writing (May 2023), ChatGPT-4 has the new feature of using external plug-ins. With the enabled plug-in of Wolfram Alpha, ChatGPT-4 can do mathematical calculation accurately and deliver the correct answer B directly for Question 263.

\subsection{Analysing medical images using visual input is impossible yet}
Despite being equipped with a new feature of visual input, ChatGPT-4 falls short in its ability to describe the content of medical images. In the evaluation of the ACR TXIT exam, when presented with an image such as the one in Question 116 as depicted in Fig.\,\ref{Fig:ExemplaryImageQuestions} via a URL link, ChatGPT-4 failed to provide any meaningful context regarding the image. Instead, it provided a generic response, stating that it could not view images as an AI language model. While ChatGPT is capable of generating descriptions of images based on accompanying text descriptions, such descriptions may not always align with the actual content of the image. The same observation is also reported in \cite{waisberg2023gpt}, where ChatGPT fails to identify retinal fundus images. As such, the current version of ChatGPT is not capable of analysing medical images in the same manner as a radiation oncologist and providing relevant diagnoses and treatment recommendations. Combining ChatGPT with medical imaging processing networks like ChatCAD \cite{wang2023chatcad} is promising in enhancing its capacity to analyze medical images in radiation oncology.

\subsection{Potential for summarizing guidelines with in-context learning}
In the TXIT evaluation, ChatGPT demonstrated its proficiency in accurately answering Question 107 when presented with a summary of the PORTEC-3 trial \cite{de2019adjuvant}. In the evaluation of Gray Zone cases, ChatGPT is also able to provide concise summaries of other experts' opinions and update its initial recommendation based on other experts' opinions. The risk of hallucination is reduced after seeing other experts' opinions. As clinical guidelines are frequently updated, many clinicians may not be familiar with the latest details. Fortunately, ChatGPT is adept at summarizing text documents and, with in-context learning techniques \cite{dong2022survey,zheng2023progressive,he2023icl}, it is capable of rapidly acquiring new knowledge. When information is presented in context, ChatGPT can provide suggestions with a high degree of confidence. Consequently, ChatGPT has the potential to considerably assist clinicians in understanding updated guidelines and providing up-to-date treatment recommendations to patients based on the latest guidelines. Due to the extensive nature of many guidelines, like the \href{https://www.nccn.org/guidelines/category_1}{NCCN cancer treatment guidelines}, which often exceed the 4k token limit of the current ChatGPT interface, access to higher token limits, such as 32k tokens, is necessary for effective utilization in such applications.

\subsection{Improvement with further domain-specific fine-tuning}
\modified{At the time of this manuscript preparation, ChatGPT is not qualified as a specialist in radiation oncology yet, since ChatGPT still lacks in-depth knowledge in many areas, as revealed by the TXIT exam and the Gray Zone cases.} Similar to other fine-tuned LLMs like Med-PaLM \cite{singhal2022large}, ChatDoctor \cite{yunxiang2023chatdoctor}, and HuaTuo \cite{wang2023huatuo}, a radiation oncology domain-specific, fine-tuned LLM can be trained to better assist radiation oncologists in decision making for real clinical cases. The vast collection of Red Journal Gray Zone cases \cite{palma2017introducing} and the latest guidelines provides ample data that can be automatically extracted for use in training such a domain-specific model, which is promising as our future work.

\subsection{Capacity extension with external plug-ins}
As a LLM, the fundamental function of ChatGPT is text generation. Hence, it has many limitations in many specialized tasks, for example, mathematical calculation in Fig.\,\ref{Fig:Q263Math}. Some of such limitations can be addressed by the new feature of external plug-ins. Especially, as the training data for ChatGPT-4 dates back to 2021 September, it has very limited knowledge on the latest updates of guidelines for radiation oncology. Enabling the internet browsing function can improve ChatGPT-4's responses to such queries.

\subsection{Growing capabilities of ChatGPT}
In this ACR TXIT exam evaluation, ChatGPT-4 has demonstrated its superiority to ChatGPT-3.5 in both the general radiation oncology field and various knowledge sub-domains, despite its slower generation speed and limited access. Notably, the questions on clinical trials suggest that ChatGPT-3.5 and ChatGPT-4 were trained on similar data sets. Therefore, the enhanced performance of ChatGPT-4 may be attributed more to its superior interpretability and generation capabilities than to the potentially increased amount of training data. In the evaluation of Gray Zone cases, ChatGPT-4 is able to update its own recommendation based on other experts' recommendations. With ongoing technical advancements, continuously expanding training data, more feedback via RLHF \cite{christiano2017deep}, and more external plug-ins, future iterations of ChatGPT are expected to deliver even more impressive performance in all medical fields, including radiation oncology.

\subsection{Limitations of this study}
The study in this work is not without its limitations. The ACR TXIT exam and the Gray Zone cases in this work represent only a narrow spectrum of knowledge in radiation oncology, as only 293 questions from the TXIT exam and 15 cases from the Gray Zone collection were evaluated. While the gaps in knowledge such as gynecology and brachytherapy were detected in this work, other benchmark tests are likely to find different deficiencies across medical domains.

Another limitation is that our performance benchmark applies to ChatGPT-3.5 and ChatGPT-4 in a specific time window (from April to August 2023). LLMs including ChatGPT are frequently updated. Therefore, the performance is highly likely to vary when new updates are applied to ChatGPT in the future.

In the work of \cite{holmes2023evaluating}, the superiority of ChatGPT over Bard in radiation oncology physics has been demonstrated. Because of its superior performance, this work focuses on the benchmark of ChatGPT's performance in a broader field of clinical radiation oncology. However, ChatGPT's performance has not been compared comprehensively with other LLMs on the TXIT exam and the Gray Zone cases in this work. In this work, the performance of LLaMA-2 \cite{touvron2023llama2} on the TXIT exam has been evaluated. It achieves an average accuracy of 34.81\% (more details in the supplementary material in our GitHub), which is lower than ChatGPT-4. LLaMA-2 with more parameters such as the 13b and 70b models are likely to have better performance. However, due to our hardware limitation, we are not able to evaluate them at the current stage. Some LLMs, in particular Med-PaLM \cite{singhal2022large} and its newer version Med-PaLM 2 \cite{singhal2023towards2}, are fine-tuned specifically for medical applications, which have the potential to outperform ChatGPT-4 in radiation oncology. Nevertheless, such comparison will be our future work once we have granted the access to Med-PaLM (version 1 or 2).

The TXIT exam has a standard answer to each question and hence it can be evaluated objectively and accurately. However, for the complex Gray Zone cases, no gold standards exist to assess the accuracy of ChatGPT-4's responses. As a consequence, intra-rater and inter-rater variability is one major limitation in our evaluation on the Gray Zone cases. To draw more definitive conclusions, a robust research design, such as providing concordance training for evaluators before the assessment, is recommended. Nonetheless, the present analysis provides valuable insights into expert perceptions of ChatGPT-4's proficiency in clinical decision-making within radiation oncology.

\section{Conclusion}
This study benchmarks the performance of ChatGPT-3.5 and ChatGPT-4 on the 38th ACR TXIT exam in radiation oncology and the 2022 Red Journal Gray Zone cases. For the TXIT exam, ChatGPT-3.5 and ChatGPT-4 have achieved accuracies of 63.65\% and 74.57\%, respectively, indicating the advantage of the latest ChatGPT-4 model. Based on the TXIT exam, ChatGPT-4's strong and weak areas in radiation oncology are identified to some extent. \modified{Specifically, ChatGPT-4 demonstrates better knowledge in statistics, CNS \& eye, pediatrics, biology, and physics than in bone \& soft tissue and gynecology, as per the ACR knowledge domain. Regarding clinical care paths, ChatGPT-4 performs better in diagnosis, prognosis, and toxicity than brachytherapy and dosimetry. And it lacks proficiency in in-depth details for clinical trials.} For the Gray Zone cases, ChatGPT-4 is able to suggest a personalized treatment approach to each case with high correctness and comprehensiveness considering the patient's historic treatment response, personal priority, and quality of life. Most importantly, it provides novel treatment aspects for many cases, which are not suggested by any human experts. Both evaluations have demonstrated the potential of ChatGPT in medical education for the general public and cancer patients, as well as the potential to aid clinical decision-making, while acknowledging its limitations in certain domains. Despite these promising results, \modified{ChatGPT-4 is not competent for clinical use yet.} ChatGPT's answers currently always have to be verified, because of the risk of hallucination, which is one of main remaining issues that will need to be addressed by future developments.

\section*{Conflict of Interest Statement}

The authors declare that the research was conducted in the absence of any commercial or financial relationships that could be construed as a potential conflict of interest.


\section*{Funding}
This research was not supported by any particular funding.

\section*{Data Availability Statement}
The datasets [GENERATED/ANALYZED] for this study can be found in the GitHub repository \href{https://github.com/YixingHuang/ChatGPT-Benchmark-on-Radiation-Oncology}{https://github.com/YixingHuang/ChatGPT-Benchmark-on-Radiation-Oncology}.



\end{document}